\documentclass[journal]{IEEEtran}
\usepackage{algorithm}  
\usepackage{algpseudocode}  
\usepackage{amsmath}

\usepackage{graphicx}
\usepackage[]{algpseudocode}
\usepackage{algorithmicx,algorithm}
\usepackage{amsmath}
\usepackage{titlesec}
\usepackage{multirow}  
\usepackage{amsthm,amsmath,amssymb}
\usepackage{graphicx}
\usepackage{float}
\usepackage{subfigure}
\usepackage{mathrsfs}
\usepackage{url}
\usepackage{array}
\newcolumntype{C}[1]{>{\centering}p{#1}}
\usepackage{makecell}
\usepackage{diagbox}
\usepackage{amsmath}
\usepackage{gensymb}
\usepackage{cite}

\ifCLASSOPTIONcompsoc
 \usepackage[caption=false,font=normalsize,labelfont=sf,textfont=sf]{subfig}
\else
 \usepackage[caption=false,font=footnotesize]{subfig}
\fi

\UseRawInputEncoding
\begin{document}
\vspace{-15mm}
\title{Electromagnetic Channel Model for Near Field MIMO Systems in The Half Space}

\author{Yuhua Jiang and Feifei Gao

\thanks{Y. Jiang, and F. Gao are with Institute for Artificial Intelligence, Tsinghua University (THUAI), 
State Key Lab of Intelligent Technologies and Systems, Tsinghua University, 
Beijing National Research Center for Information Science and Technology (BNRist), Beijing, P.R. China (email:  jiangyh20@mails.tsinghua.edu.cn, feifeigao@ieee.org).
}
}

\maketitle
\vspace{-15mm}
\begin{abstract}
In most multiple-input multiple-output (MIMO) communication systems, the amount of information that can be transmitted reliably depends on the effective degrees of freedom (EDoF) of the wireless channel.
Conventionally, one can model the channel matrix and study the EDoF, based on an electromagnetic (EM) channel model that is built with the free space Green’s function.
However, the EDoF of free-space channel model may not fit the practical scenario when EM waves only transmit above the ground. 
In this paper, we analyze the EDoF for both discrete and continuous aperture MIMO systems in the half space.
We also propose an approach to quickly calculate the Green function in the half space from the Sommerfeld identity.
Simulation results show that the difference between the EDoF in the half space and that in the free space is non-negligible for the near field communications, which indicates that the ground exerts noticeable influence on EDoF. The proposed study establishes a fundamental electromagnetic framework for MIMO wireless communication in the half space.
\end{abstract}

\IEEEpeerreviewmaketitle

\section{Introduction}
Multiple-input-multiple-output (MIMO) technology using spatial
multiplexing has been developed to enhance the channel
capacity of modern wireless communications \cite{22}. Different electromagnetic (EM) modes, as orthogonal bases, have been employed in MIMO systems.
Typically, the channel matrix is modeled
based on the scalar Green’s function for MIMO systems in
free space \cite{7}.  

The degrees of freedom (DoF) of the MIMO system has been investigated from EM
perspectives.  In \cite{8}, the authors derive the explicit results for the communications modes between rectangular volumes and between small volumes. 
Recently, the number of single input single output (SISO) subchannels defined as the effective degrees of freedom (EDoF) has aroused much interest, because the EDoF fixes the maximum achievable capacity in the MIMO system \cite{10}. Meanwhile, the upper bound of the EDoF is used to calculate the
DoF \cite{12}.
The evaluation of the EDoF is a complicated problem, which can be studied based on two kinds of models, i.e., the conventional channel model and the EM model.

Based on the conventional channel model of the MIMO system in the free space, \cite{18} calculates EDoF by studying the spectral efficiency of MIMO systems with different shapes of arrays.  In \cite{19}, the authors estimate the EDoF of the MIMO system when the channel matrix is subject to empirical distribution. However, the conventional channel model is less accurate than the EM model of the MIMO system.

Based on the EM model of the MIMO system in
free space, \cite{16} discusses the method to calculate EDoF, the limit
of EDoF and the optimal number of sources and receivers. Furthermore, an EM model for the hologram
MIMO has been proposed in \cite{11}, where the channel
matrix is built with plane-wave expansion.
The EM model is useful for linking EDoF with the capacity limit of the MIMO system. 

However, the channel models in [5]-[8] are all built in the free space scenario, while in the near field, the free space does not fit the reality well because the EM waves can only transmit above the ground \cite{14}. Hence, the boundary condition on the ground should be considered to generate the more accurate channel model, namely, more accurate channel model should be established in the half space, which could help to predict the channel capacity more precisely \cite{16}.

In this paper, we analyze the EDoF for both discrete and continuous aperture MIMO systems in the half space.
We also propose an approach to quickly calculate the Green function in the half space from the Sommerfeld identity.
Simulation results show that the difference between the EDoF in the half space and that in the free space is significant for the near field communications, which indicates that the ground exerts noticeable influence on the channel capacity. With the increase of the antenna number, the EDoF for discrete MIMO systems converges to that for continuous MIMO systems.


\section{Discrete-aperture MIMO System}
Suppose the source is equipped with $N$ antennas and the receiver is equipped with $M$ antennas.
The channel vector for the $m$th antenna of the receiver is defined as $\mathbf{h}_m$ $(m=1,\cdots,M)$, which can be written as \cite{16}
\begin{align}
\mathbf{h}_m=
\left[G_{m1} \cdots,G_{mN} \right]^T,
\end{align}
where $G_{mn}$ is the scalar Green’s function linking the $m$th transmit antenna and the $n$th receiver antenna.
The overall channel $\mathbf{H}$ for the receiver is obtained by 
merging $\mathbf{h}_m$ $(m=1,\cdots,M)$ as
\begin{equation}
\mathbf{H}=\left[\mathbf{h}_1,\cdots,\mathbf{h}_M\right]^T.
\end{equation}
Define the correlation matrix as $\mathbf{R}=\mathbf{H}^H\mathbf{H}$. The EDoF of the discrete-aperture MIMO system is a function of $\mathbf{R}$ represented as $\Xi(\mathbf{R})$, and can be approximately calculated as \cite{1}, \cite{2}
\begin{equation}
\Xi(\mathbf{R})=\left(\frac{\operatorname{tr}(\mathbf{R})}{\|\mathbf{R}\|_{F}}\right)^{2}=\frac{\left(\sum_{i} \sigma_{i}\right)^{2}}{\sum_{i} \sigma_{i}^{2}},
\label{dis}
\end{equation}
where $\sigma_{i}$ is the $i$th eigenvalue of $\mathbf{R}$. 
In the far-field communications, the leading eigenvalue of $\mathbf{R}$ is significantly larger than the other eigenvalues, and the EDoF is close to one corresponding to the only communication mode where a plane wave travels from the transmitting to the receiving antenna \cite{9}.

\section{Continuous-aperture MIMO System}
Recently, the study of continuous-aperture MIMO systems has aroused much interest, because the continuous-aperture MIMO systems can sufficiently exploit physical properties of spatial electromagnetic waves, leading to extreme spatial resolution, high
spectrum efficiency, and high energy efficiency \cite{20}. 
The formulation of the EDoF in discrete-aperture MIMO systems can be extended to continuous-aperture MIMO systems with the help of auto-correlation kernel function. 

Suppose  the continuous-aperture source and the continuous-aperture receiver are uniform linear arrays (ULA) whose lengths are $L_s$ and $L_r$, respectively. Let $S$ denote the region of the source and $R$ denote the region of the receiver.
Define the Green function relating two arbitrary locations $\textbf{r}_s \in S,\textbf{r}_r \in R$ as $G(\textbf{r}_r,\textbf{r}_s)$, and define the auto-correlation kernel $K(\textbf{r}_s,\textbf{r}_s')$ that correlates two locations in the source region $\textbf{r}_s \in S,\textbf{r}_s' \in S$
as
\begin{equation}
K(\textbf{r}_s,\textbf{r}_s')=\int_R G^H(\textbf{r}_r,\textbf{r}_s) G(\textbf{r}_r,\textbf{r}_s') d\textbf{r}_r.
\end{equation}

The channel correlation matrix $\textbf{R}$ is then derived from $\mathbf{R}=\mathbf{H}^H\mathbf{H}$
under the condition  $M\rightarrow \infty,N\rightarrow \infty$ while the sizes of the source and the receiver are fixed. 
The $(n_1,n_2)$th element of $\mathbf{R}$  corresponds to the  channel correlation between the $n_1$th and $n_2$th antennas of the source, and has the asymptotic representation:
\begin{align}
\mathbf{R}_{(n_1,n_2)}=|\sum_{m=1}^M G_{mn_1}^H G_{mn_2}|^2 \rightarrow \frac{M^2}{L_r^2}|K(\textbf{r}_s,\textbf{r}_s')|^2 .
\end{align}
Then the following equations hold:
\begin{align}
\|\mathbf{R}\|_{F}^{2}&= \sum_{n_1=1}^N \sum_{n_2=1}^N |\sum_{m=1}^M G_{mn_1}^H G_{mn_2}|^2\nonumber\\
&\rightarrow\frac{N^2}{L_s^2}\int_S \int_S \frac{M^2}{L_r^2}|K(\textbf{r}_s,\textbf{r}_s')|^2 d\textbf{r}_s d\textbf{r}_s'\\
\operatorname{tr}(\textbf{R})&=\sum_{m=1}^M \sum_{n=1}^N |G_{mn}|^2
\rightarrow\frac{MN}{L_s L_r}\int_S \int_R |G(\textbf{r}_r,\textbf{r}_s)|^2 d\textbf{r}_r d\textbf{r}_s,
\end{align}
where the  asymptotic representation comes from the fact:
$d\textbf{r}_s \sim \frac{L_s}{N}$, $d\textbf{r}_s' \sim \frac{L_s}{N}$, and $d\textbf{r}_r \sim \frac{L_r}{M}$.
Since $\|\mathbf{R}\|_{F}^{2}$ is $\mathcal{O}(M^2 N^2)$
and $\operatorname{tr}(\textbf{R})$ is $\mathcal{O}(M N)$, $\Xi(\mathbf{R})=\left(\frac{\operatorname{tr}(\mathbf{R})}{\|\mathbf{R}\|_{F}}\right)^{2}$ converges when $M\rightarrow \infty,N\rightarrow \infty$.
The EDoF of the continuous-aperture MIMO system is then computed as
\begin{align}
\operatorname{L}(\textbf{R})=\lim_{M,N\rightarrow \infty}\Xi(\mathbf{R})&=\lim_{M,N\rightarrow \infty}\left(\frac{\operatorname{tr}(\mathbf{R})}{\|\mathbf{R}\|_{F}}\right)^{2}\nonumber\\
&=\frac{(\int_S \int_R |G(\textbf{r}_r,\textbf{r}_s)|^2 d\textbf{r}_r d\textbf{r}_s)^2 }{\int_S \int_S |K(\textbf{r}_s,\textbf{r}_s')|^2 d\textbf{r}_s d\textbf{r}_s'}.
\label{con}
\end{align}

\section{Green Function in The Half Space}
Note that $\operatorname{L}(\textbf{R})$ in (\ref{dis}) as well as $\Xi(\mathbf{R})$ in (\ref{con}) relies on the Green function, which will be computed in this section.
\begin{figure}[t]
  \centering
  \centerline{\includegraphics[width=6.7cm,height=5cm]{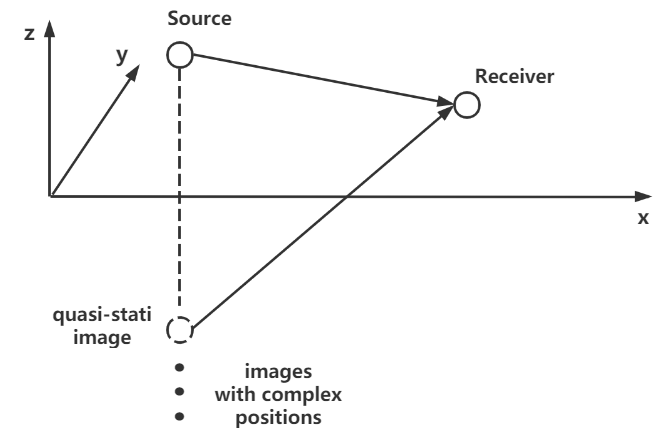}}
  \caption{The ground is in the $xy$ plane where $z=0$. 
  The images of the source are composed of a single quasi-static image and several images with complex positions.}
  \label{image}
\end{figure}
\subsection{The Closed Form Representation of The Green Function}
Since the EM waves can only transmit above the ground, the boundary condition on the ground where $z=0$ should be considered when calculating the EM field, which is written in the coordinate in Fig.~\ref{image} as
\begin{align}
\frac{\partial \phi}{\partial z}+i k_0 \beta \phi=0, \quad z=0,
\end{align}
where $\phi$ is the electric field, $k_0$ is the wave number, and $\beta$ is the normalized admittance of the ground. Then the reflection coefficient for the ground is defined as 
\begin{align}
\tilde{C}(k)=\frac{k_{z}-k_{0} \beta}{k_{z}+k_{0} \beta},
\label{C}
\end{align}
where $k_z$ is the wave number in $z$ direction with the form
\begin{align}
k_z=\sqrt{k_0^2-k^2}.
\end{align}

In the half space, the exact expression of the Green's function for the Helmholtz equation relating $\mathbf{r}_r=[x_r,y_r,z_r]^T$ and $\mathbf{r}_s=[x_s,y_s,z_s]^T$ 
can be represented by the method of images as \cite{21}
\begin{align}
G\left(\mathbf{r}_r, \mathbf{r}_s\right)=& \frac{i}{4 \pi} \int_{0}^{\infty} \frac{e^{i k_{z}\left|z_r-z_{s}\right|}k J_{0}(k \rho)}{k_{z}} dk \nonumber\\
&+\frac{i}{4 \pi} \int_{0}^{\infty} \frac{e^{jk z_{z}\left(z_r+z_{s}\right)} k J_{0}(k \rho)}{k_{z}} \tilde{C}(k)dk,
\label{2}
\end{align}
where the horizontal distance in the $xy$ plane is $\rho=\left[\left(x_r-x_{s}\right)^{2}+\left(y_r-y_{s}\right)^{2}\right]^{1 / 2}$ and $J_{0}$ is the $0$th-order Bessel function of the first kind. The integral in (\ref{2}) is referred to as Sommerfeld integral whose analytical solution is not discovered so far \cite{5271026}.
To simplify the numerical evaluation of the Sommerfeld integral, we can use the Sommerfeld identity:
\begin{align}
\frac{e^{i k R}}{R}=i \int_{0}^{\infty} J_{0}\left(k \rho\right) e^{i k_{z}\left(z+z^{\prime}\right)} \frac{k}{k_{z}} d k,
\label{11}
\end{align}
where $R=\sqrt{\rho^{2}+\left(z+z^{\prime}\right)^{2}}$. 
Note that when $k_{0}=0$, $\tilde{C}(k)$ is constantly equal to $1$. The image corresponding to $\tilde{C}(k)=1$ is named as the quasi-static image.
Using the Sommerfeld identity to extract the quasi-static term of the image, we can rewrite (\ref{2}) as
\begin{align}
G\left(\mathbf{r}, \mathbf{r}_{s}\right)=& \frac{\exp \left(ik_{0} D_{1}\right)}{4 \pi D_{1}}+\frac{\exp \left(ik_{0} D_{2}\right)}{4 \pi D_{2}} \nonumber\\
&+\frac{i}{4 \pi} \int_{0}^{\infty} \frac{e^{i k_{z}\left(z_r+z_{s}\right)} kJ(k \rho)}{k_{z}}\left(\tilde{C}(k)-1\right) dk,
\label{4}
\end{align}
where $D_{1}=\left[\rho^{2}+\left(z_r-z_{s}\right)^{2}\right]^{1 / 2}$, $D_{2}=\left[\rho^{2}+\left(z_r+z_{s}\right)^{2}\right]^{1 / 2}$. 

The last integral in (\ref{4}) is hard to calculate straightforwardly due to the oscillation of the integrand. Thus, we apply the exponential expansion method.
The function $\tilde{C}(k)-1$ can be exponentially expanded with high accuracy in regard of $k_z$ as \cite{sound}
\begin{align}
\tilde{C}(k)-1 \approx \sum_{n=1}^{Q} a_{n} e^{-b_{n} k_z},
\label{3}
\end{align}
where $a_{n}$ and $b_{n}$ are complex numbers and $Q$ is an integer. 
Utilizing (\ref{3}) to simplify (\ref{4}), we obtain a closed-form approximation of $G\left(\mathbf{r}_r, \mathbf{r}_{s}\right)$ as 
\begin{align}
G\left(\mathbf{r}_r, \mathbf{r}_{s}\right)\approx& \frac{\exp \left(ik_{0} D_{1}\right)}{4 \pi D_{1}}+ \frac{\exp \left(ik_{0} D_{2}\right)}{4 \pi D_{2}} \nonumber \\
&+\sum_{n=1}^{Q} a_{n} \frac{\exp \left(ik_{0} R_{n}\right)}{4 \pi R_{n}},
\label{5}
\end{align}
where $R_{n}$ is a complex distance with positive real part and is given by
\begin{align}
R_{n}^{2}=\rho^{2}+\left(z_r+z_{s}+b_{n} i\right)^{2}.
\end{align}
The three terms in (\ref{5}) are regarded as the contributions from the original source, its quasi-static image, and several images with complex positions, respectively.
Note that (\ref{5}) applies for sources and receivers located anywhere within the half-space above the ground, for a single set of constants $a_{n}$ and $b_{n}$. The Green function in the free space is the first term in (\ref{5}), i.e.,
\begin{align}
G_{free}\left(\mathbf{r}_r, \mathbf{r}_{s}\right)= \frac{\exp \left(ik_{0} D_{1}\right)}{4 \pi D_{1}}.
\end{align}

\subsection{Obtaining The Coefficients $a_{n}$ and $b_{n}$}
The closed form expression for the Green's function is quite simple. However, obtaining the coefficients $a_{n}$ and $b_{n}$ for (\ref{5}) is not straightforward. Note that the integrand in (\ref{4}) has a pole as well as a branch point at $k=k_{0}$. To avoid the rapid variation near the branch point, we select a deformed path of integral in (\ref{4}), defined as \cite{5}
\begin{align}
k_{z}=k_{0}\left[i \xi+\left(1-\frac{\xi}{T}\right)\right], \quad 0 \leqslant \xi \leqslant T,
\label{deform}
\end{align}
where $T$ is adjustable and controls the real axis intercept of the path. For near-field computation, choosing $T=10$ can ensure high accuracy of the approximation in (\ref{5}), as is proven in \cite{sound2}.

On the deformed path in (\ref{deform}), $\tilde{C}(k)-1$ can be approximated by an exponential expansion in regard of $\xi$ as
\begin{align}
\tilde{C}(k)-1 \approx K(\xi) &\overset{\Delta}{=} \sum_{n=1}^{Q} A_{n} e^{B_{n} \xi} \nonumber\\
&=\sum_{n=1}^{Q} A_{n} \exp \left(B_{n} \frac{\left(k_{0}-k_{z}\right) T}{k_{0}(1-i T)}\right) .
\end{align}
Comparing (19) and (24), $a_{n}$ and $b_{n}$ can be written in terms of $A_{n}$ and $B_{n}$ as
\begin{align}
a_{n}=A_{n} \exp \left(\frac{B_{n} T}{1-i T}\right), \quad b_{n}=B_{n} \frac{T}{k_{0}(1-i T)}.
\end{align}
Thus, we translate the task of computing $a_{n}$ and $b_{n}$ into the task of computing $A_{n}$ and $B_{n}$, which will be discussed in the next subsection.

\subsection{The Sampling Method to Obtain $A_{n}$ And $B_{n}$}
In order to obtain $A_{n}$ and $B_{n}$ such that $\tilde{C}(k)-1$ can be approximated by $K(\xi)$, we apply the modified Prony method.

Define the polynomial $f(x)$ with roots $\zeta_n=\exp{(\frac{B_n T}{W})}$ as
\begin{align}
f(x)=\prod_{n=1}^{Q}(x-\exp{(\frac{B_n T}{W})})=x^Q+C_{Q-1}x^{Q-1}+\cdots+C_{0}.
\label{A4}
\end{align}
Thus, $y_n(w)=\exp{(\frac{B_n T w}{W})}$ ($n=1,\cdots,Q$) satisfies the $Q$th-order linear difference equation with the characteristic equation as $f(x)$:
 \begin{align}
y_n(w+Q)+& C_{Q-1} y_n(w+Q-1)&+\cdots+C_{0} y_n(w)=0 \nonumber\\ 
&\quad w=1,\cdots,W-Q
\end{align}
Define the uniformly sampling points $F(w)=K(\frac{T w}{W})=\sum_{n=1}^{Q} A_{n} \exp{(\frac{B_n T w}{W})} (w=1,\cdots,W)$. Since each $F(w)$ is the linear combination of $y_n(w)$, $n=1,\cdots,Q$, $F(w)$ satisfies the difference equation of the same form. Define
\begin{align}
&\textbf{A}=\left[\begin{array}{cccc}
F(1) & F(2) & \cdots & F(Q) \\
F(2) & F(3) & \cdots & F(Q+1) \\
\vdots & \vdots & \ddots & \vdots \\
F(W-Q) & F(W-Q+1) & \cdots & F(W-1)
\end{array}\right]\\
&\mathbf{g}=\left[\begin{array}{c}
C_0 \\
C_1 \\
\vdots \\
C_{Q-1}
\end{array}\right] \quad
\mathbf{b}=-\left[\begin{array}{c}
F(Q+1) \\
F(Q+2) \\
\vdots \\
F(W)
\end{array}\right].
\end{align}
Then, there is
\begin{align}
\mathbf{A} \mathbf{g}=\mathbf{b}.
\end{align}
After obtaining the values of $F(w)$ from the original expression $\tilde{C}(k)-1$ by using (\ref{C}), we can determine
$C_n (n = 0, 1,\cdots, Q - 1)$ by computing $\mathbf{g}=\mathbf{A}^\dagger \mathbf{b}$ where $\mathbf{A}^\dagger$ is the pseudo-inverse of $\mathbf{A}$.

After obtaining $C_n$, we can put them into 
the characteristic polynomial equation of (\ref{A4}) and
determine the roots $\zeta_n=\exp{(\frac{B_n T}{W})}$, which in turn yields the exponents $B_n$. Solving the linear equations $F(w)=\sum_{n=1}^{Q} A_{n} \exp{(\frac{B_n T w}{W})} (w=1,\cdots,W)$ by the least square method, we can get the value of $A_{n}$.

\begin{figure}[t]
  \centering
  \centerline{\includegraphics[width=6.7cm,height=6cm]{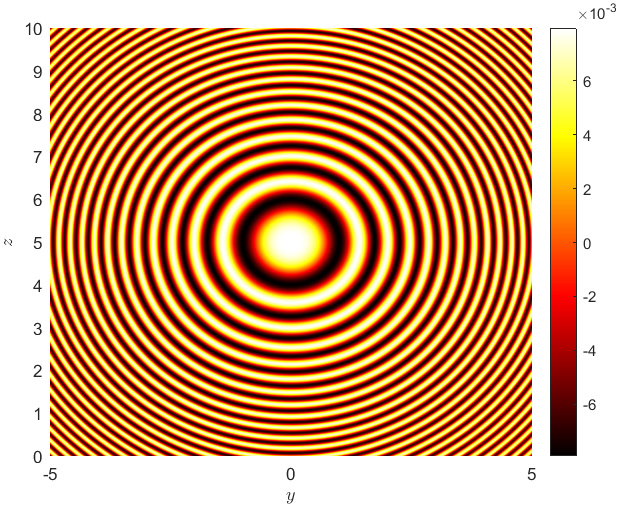}}
  \caption{The real part of the free-space Green’s function}
  \label{free}
\end{figure}

\begin{figure}[t]
  \centering
  \centerline{\includegraphics[width=6.7cm,height=6cm]{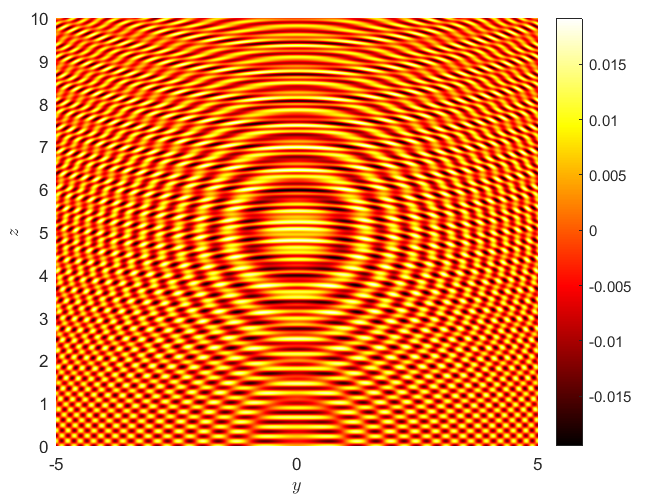}}
  \caption{The real part of the half-space Green’s function}
  \label{half}
\end{figure}

To provide a vivid picture on how the boundary condition of the ground influences the Green function,
Fig.~\ref{free} and Fig.~\ref{half} plot  the real part of the free-space
Green function and the half-space Green function in the plane of $x=10$m, for a point source located at $(0,0,5)$m. Because the influence of the ground, up-down asymmetry is shown in Fig.~\ref{half}. While in Fig.~\ref{free}, the influence of the ground is neglected and the  Green function is axisymmetric.

\section{Simulation Results and Analysis}
In the simulations, the normalized surface impedance value is chosen to be $\eta=0.3-0.1 i$.\footnote{This value is obtained from the gray clay loam of San Antonio.} Thus we have $\beta=1/\eta=3+1i$. We choose $T=10$, $Q=5$, and $W=10$ to calculate Green function in the half space, which is proved to be accurate in near field scenario \cite{sound2}.
We set the wave length $\lambda=0.1$m and define the height of the source and the receiver as $z_s$ and $z_r$ respectively.
The source and the receiver are both equipped with extremely long ULAs which are parallel to each other with $L_r=4$m and $L_s=12$m. The great lengths of ULAs are utilized to prevent communication happening in the far-field regime where the EDoF is constantly close to one \cite{9}.


\subsection{The EDoF versus The Number of Antennas}
\begin{figure}[t]
  \centering
  \centerline{\includegraphics[width=8.4cm,height=6cm]{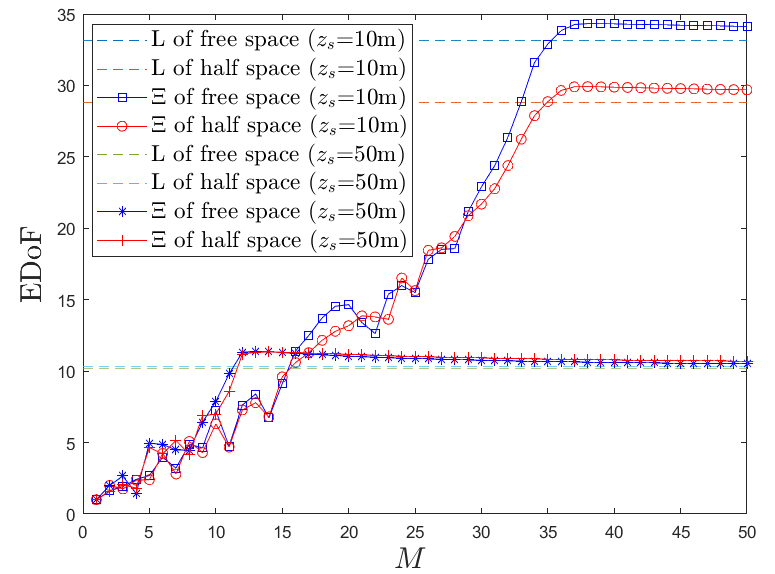}}
  \caption{The EDoF versus the number of antennas $M$.}
  \label{P2}
\end{figure}
For the discrete aperture, we suppose the numbers of the antennas at the source side and at the receiver side are the same, i.e., we set $M=N$. 
We set $z_r=1$m and explore the change of EDoF with the increase of $M$ when $\rho=10$m as is shown in Fig.~\ref{P2}. 
It is seen that, $\Xi$ increases with fluctuation until a certain number, which is defined as the \textit{optimal number} for MIMO systems, and $\Xi$ slowly decreases afterwards. 
Moreover, $\Xi$ converges to $L$ with the increase of $M$, which verifies the convergence analysis in Section~\uppercase\expandafter{\romannumeral3}.
Both $\Xi$ and $L$ in the half space are different than $\Xi$ and $L$ in the free space. While the \textit{optimal numbers} for MIMO systems are the same both in the half space and in the free space.
The difference between half space and free space when $z_s=10$m is more significant than the difference when $z_s=50$m, because when the source is more close to the ground, the influence on the EM field by the images is more remarkable.
\subsection{The EDoF versus The Horizontal Distance}
\begin{figure}[t]
  \centering
  \centerline{\includegraphics[width=8.4cm,height=6cm]{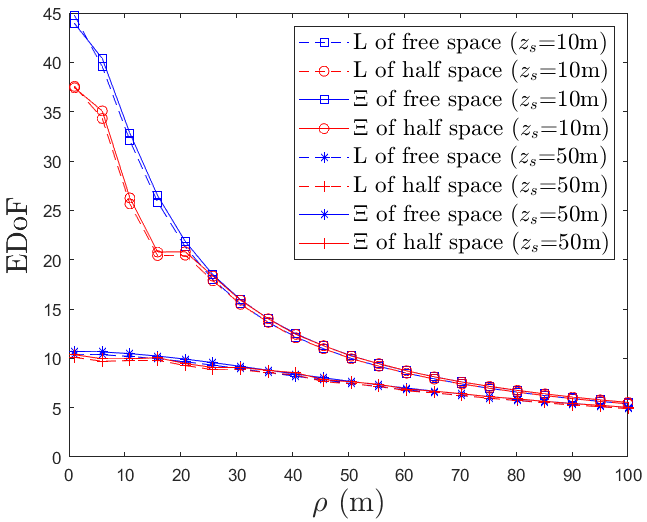}}
  \caption{EDoF versus the horizontal distance $\rho$.}
  \label{P3}
\end{figure} 
We set $z_r=1$m, $M=N=50$ and explore the change of EDoF with the increase of $\rho$ as is shown in Fig.~\ref{P3}.
It is seen that, EDoF drops with the increase of $\rho$ for both the continuous aperture and the discrete aperture.
The EDoF in the half space is less than that in the free space when $\rho$ is less than $30$m and the difference between them vanishes when $\rho$ is larger than $30$m, which indicates that the ground exerts noticeable influence on the EDoF especially in the near field.


 \subsection{The EDoF versus The Height of The Receiver}
\begin{figure}[t]
  \centering
  \centerline{\includegraphics[width=8.4cm,height=6cm]{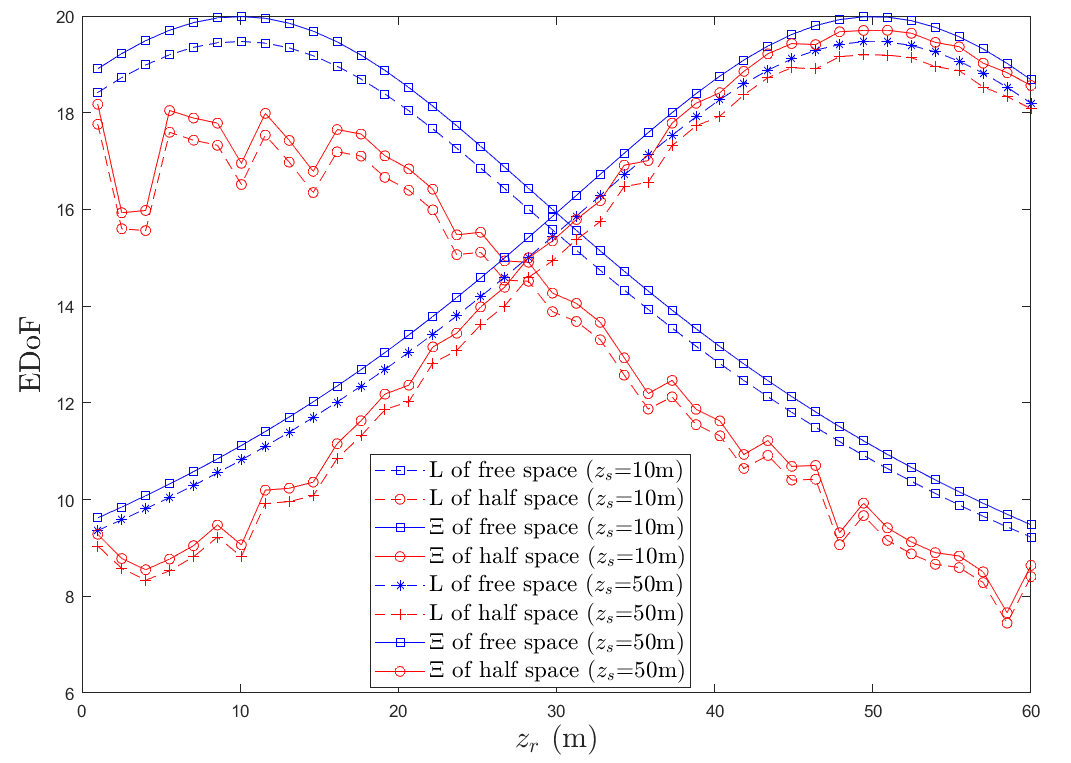}}
  \caption{EDoF versus the horizontal distance $z_r$.}
  \label{P4}
\end{figure} 
We set $\rho=25$m, $M=N=50$ and explore the change of EDoF with the increase of $z_r$ as is shown in Fig.~\ref{P4}.
In the free space, EDoF is positively related to the absolute difference of $z_r$ and $z_s$ for both the continuous aperture and the discrete aperture. In the half space, EDoF fluctuates with the change of $z_r$, and is less than the EDoF in the free space. Their difference is smallest when $z_r$ and $z_s$ are both large and the difference is largest when $z_r$ and $z_s$ are both small. This indicates that the influence of the ground diminishes when the source or the receiver moves far away from the ground.

\section{Conclusion}
In this paper, we analyze the EDoF for MIMO systems in the half space by the method to quickly calculate the Green function in the half space.
Simulation results show that the difference between the EDoF in the half space and that in the free space is pronounced for the near field communications, which indicates that the ground exerts significant influence on the EDoF. 

 \small 
 \bibliographystyle{ieeetr}
 \bibliography{IEEEabrv,my_oric1}


\ifCLASSOPTIONcaptionsoff
  \newpage
\fi



%

\end{document}